# MULTI BUNCH GMMA RAY GENERATION EXPERIMENT AT ATF[*]

TOHRU TAKAHASHI[†]

*Graduate School of Advanced Sciences of Matter Hiroshima University, 1-3-1Kagamiya Higashi-Hiroshima, 739-8530, Japan*

We construct a new detector to monitor γ yields in bunch by bunch basis for the Laser Compton experiment at the KEK ATF which is capable to separate γ rays in 5.6ns-spacing multi-bunch operation of the KEK ATF. In this article we report a result of measurement of multi bunch γ ray detection for the first time at the KEK ATF.

**1 Introduction**

The polarized beams are powerful tool for physics at the International Linear Collider (ILC) since we can choose helicity Eigen states for the initial election-positron system. The ILC is designed to provide polarized electron beams using a DC gun with GaAs photo-cathode. In addition to the polarized electron, it is highly desirable to use polarized positron beams since it increase the effective polarization of the ILC[1], where effective polarization is defined as;

$$P_{eff} = \frac{P_e - P_p}{1 - P_e P_p}$$

For example, 80% polarization of the electron and 60% positron polarization yield the effective polarization of 95%. The ILC is planned to construct the positron source by the undulator scheme. In the undulator scheme, polarized γ rays of O(10MeV) are generated by feeding high energy electron beams in the main LINAC into a more than 150m long helical undulator, then γ rays are impinged into the metal target to generate positrons[2]. An optional scheme to make polarized γ rays is the Compton scheme. In Compton scheme, γ rays are generated via Compton scattering of laser photons off electrons. The Compton scheme has several advantages such as; the electron energy necessary for O(10MeV) photons is O(1GeV) which is much lower than those for the undulator scheme, the polarization of γ rays are easily controlled by the laser polarization. The principle of the Laser Compton scheme has been demonstrated by previous works [3]. While it is attractive for the polarized positron sources for the ILC , a price to pay for the Compton scheme is the intensity of the γ rays. We need high intensity and with high repetition laser pulses such as O(100mJ/pulse) with O(100MHz) repetition. We plan to achieve the required performance using a laser pulse stacking cavity in which laser pulses are accumulated coherently and its energy are enhanced inside the cavity. We have developed and tested prototype cavity at the KEK-ATF as was reported in [4].

---

[*] This work is supported by "Grant-in-Aid for Creative Scientific Research of JSPS (KAKENHI 17GS0210)" project of the Ministry of Education, Science, Sports, Culture and Technology of Japan (MEXT) and Quantum Beam Technology Program of JST.
[†] Collaborators: T.Akagi, M.Kuriki, R.Tanaka, H.Yoshitama(Hiroshima University), S.Araki, Y.Funahashi, Y.Honda, T.Omori, T.Okugi, H.Shimizu, N.Terunuma, J.Urakawa(KEK, High Energy Accelerator Rearch Organization) H.Kataoka, T.Kon(Seikei University), K.Sakaue, M.Washio(Waseda University) and French-Japan Collaboration.

One of the important issues for the ILC positron source is to generate multi-bunch positrons with stable intensity. The temporal bunch separation of the KEK-ATF is 5.6 ns and it is operative for 1 to 10 bunches per train, however, in the previous experiment, the γ ray detector equipped with CsI crystal did not have capability to observe γ rays from each bunch separately, because the decay constant of fluorescence is about 20ns even after filtering out faster component. Therefore we improved detection system to separate γ rays from laser-electron scattering which happen every 5.6 ns. In this article, we report preliminary results of multi-bunch photon detection from the laser Compton experiment for the first time at the KEK-ATF .

**2 Experimental Setup**

*2.1. Optical Resonant Cavity*

The schematic of the optical resonant cavity and laser optics are shown in Figure 1.

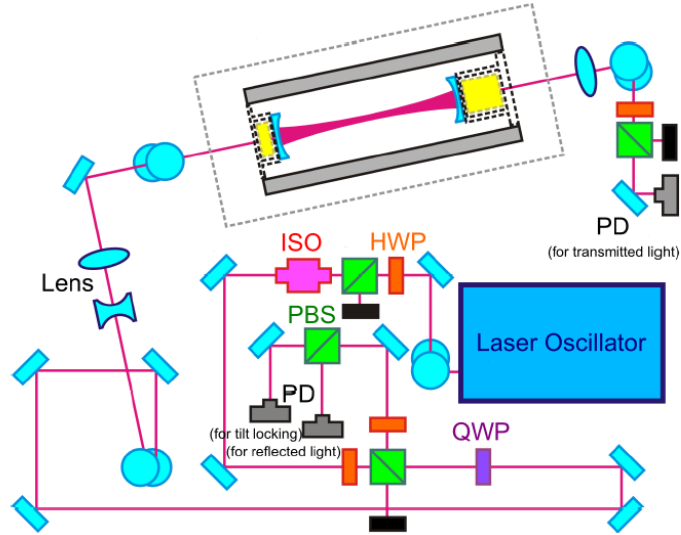

Figure 1. Schematic of the optical system for the optical resonant cavity.

We used a Mode Locked pulsed laser (COURGAR, Time-Bandwidth Products). The wave length of the laser is 1064 nm. The pulse energy is 25 nW with the repetition of 357 MHz. The optical resonant cavity is a 2 mirror Fabry-Perot type. The finesse of the cavity was measured to be 1800 which corresponds to the power enhancement factor of about 600. It was consistent with expectation from the reflectivity of mirrors of the cavity. A feedback system was constructed to keep the optical cavity on resonance with the laser pulse as well as to synchronize laser pulses with the electron bunches in the ATF. The schematic o the feedback system is shown in Figure 2.

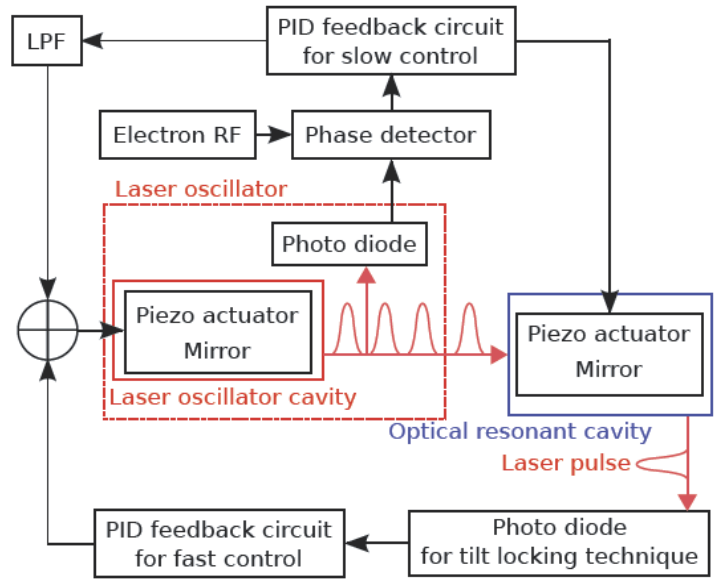

Figure 1.Diagram of the feedback and timing synchronization system.

The laser spot size inside the cavity at the laser-electron interaction point was about 30 μm (1 σ) which was estimated from the measured angular divergence the laser pulses outside the cavity. The laser power in the cavity was calculated from the measured transmitted power outside of the cavity. It was monitored during the experiment and was about 1.5kW during the experiment.

*2.2. Detector and Data Acquisition system*

To detect γ rays in bunch by bunch basis, we installed new detector consisting of a BaF2 scintillator and a Photo-Multiplier Tube (PMT) R3377 (Hamamatsu Photonics). Because the BaF2 crystal has two scintillation components with different wave length and decay constant, 310nm/600ns and 220nm/0.8ns, an optical filter was inserted between the crystal and the PMT to eliminate the slower component. The rise time of the PMT is 0.7 ns when operating at nominal high voltage. Prior to the beam experiment, detector was calibrated using cosmic rays. According to expected number of γ rays in the Compton experiment, the detector would be saturated if it was operated with the nominal voltage. Since the time response of the detector could be deteriorated for lower voltage, we took a cosmic ray data for pulse height as well as rise time with various high voltages and determined operation voltage during the beam experiment.

The waveform from the PMT for each laser electron beam interaction was recorded by a digital oscilloscope DSO5054a (Agilent Technologies) which has the bandwidth of 500MHz and the maximum sampling frequency 4GHz. The oscilloscope was placed close to the detector to avoid attenuation of higher frequency component in the signal. In parallel to the waveform, detector signal was sent to the ADC and total charge from a pulse train was recorded. The data acquisition system was triggered by the clock signal from the ATF accelerator which is synchronized with the ATF beam revolution.

### 3. Result

A typical waveform recorded in the experiment is shown in Figure 3 which was recorded during 10 bunch operation of the ATF.

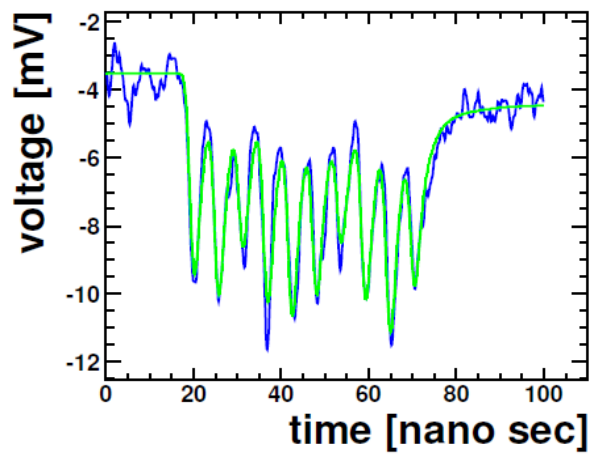

Figure 3. A typical waveform for 10 bunch data. The blue line is the raw detector signal. A result of a fitting taking the fluorescence and the detector response into account is overlaid by a green smooth line.

We clearly observed spikes of 5.6 ns separation in the waveform which correspond to γ rays from each bunch in a train. A broad shift of the baseline can be understood as a contribution of leakage of the slower component of fluorescence from the optical filter.

To estimate number of photons we fitted the waveform with a function which takes into account time dependence of the fluorescence in the BaF2 crystal, light correction in the crystal and detector response. A result of fitting is superimposed on the waveform in Figure 3. The beam current for each train in the ATF was measured the DC Current Transformer (DCCT) while relative intensity of between electron bunches was monitored by the Wall Current Monitor (WCM).

The number of γ rays for bunch by bunch basis normalized the laser power and the electron intensity is shown in Figure 6.

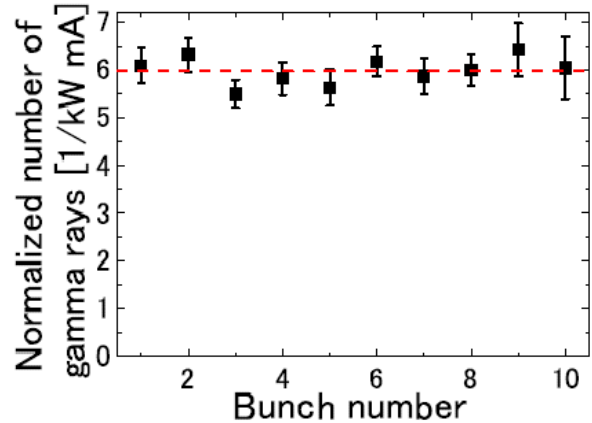

Figure 4. Normalized γ ray yields from each bunch for the 10 bunch operation. The dashed line is an expectation from numerical calculation

The expected γ ray yields was estimated by CAIN [5] program with the laser and the electron beam parameters. The γ ray yields were consistent with the expectation within the error and no bunch dependence for γ ray yield was observed.

**4. Summary**

The detector system of the laser Compton experiment at the KEK-ATF was improved to separate γ rays from each electron bunch in multi-bunch operation. The γ ray yields were observed with bunch by bunch basis up to 10 bunches/train. Observed γ ray yields were consistent with numerical expectation and no bunch dependence of the yield was observed. It was an encouraging step for the positron source R&D. Several works are in progress to improve performance of the detector and the data acquisition system toward future experiments;

- New optical filter will be inserted between the BaF2 crystal and the photo cathode of the PMT to cut leakage of slower component from BaF2 crystal.
- The signal output circuit of the PMT will be modified to avoid saturation for more intense γ ray yield which will be expected in future experiment.
- New data acquisition system is under development to record the data for every beam revolution in the ATF.

**Acknowledgments**

Data shown in this article was taken before the Great East Japan Earthquake. The KEK-ATF is back after the Great East Japan Earthquake in June 2011 after the earthquake and precise alignment is in progress for full recovery. The authors would like to thank all ILC colleagues for their warm messages and supports. Particularly we would like to thank members for ILC source working group for valuable discussions

**References**

1. G. Moortgat-Pick et al., Phys. Rep 460 (2008) 131.H. Müller and B. D. Serot, Phys. Rev. C52, 2072 (1995).
2. ILC-REPORT-2007-001(2007)
3. M. Fukuda et al., Phys. Rev. Letts 91 (2003) 164801.T. Omori et al., Phys. Rev. Letts 96 (2006) 114801.
4. For comprehensive summary of previous works, see S. Miyoshi, Ph D. thesis,
5. ADSM Hiroshima University, 2010.
6. http://www.huhep.org/Home/thesis/doctor/2011miyoshi.pdf
7. http://lcdev.kek.jp/~yokoya/CAIN/cain235/